\documentclass[preprint,floats,aps,epsfig,nofootinbib,amssymb]{revtex4}
\usepackage{mathrsfs}
\usepackage{slashed}
\usepackage{graphicx,color}
\usepackage{epsfig}
\usepackage{subfigure}
\usepackage{epsfig}
\usepackage{dcolumn}
\usepackage{bm}

\begin{document}

\title{Geometrical Model for Non-Zero $\theta_{13}$}

\author{\bf Jun-Mou Chen, Bin Wang and Xue-Qian Li}

\affiliation{School of Physics, Nankai University, Tianjin, 300071,
China}

\begin{abstract}
Based on Friedberg and Lee's geometric picture by which the
tribimaximal Pontecorvo-Maki-Nakawaga-Sakata leptonic mixing matrix
is constructed, namely, corresponding mixing angles correspond to
the geometric angles among the sides of a cube. We suggest that the
three realistic  mixing angles, which slightly deviate from the
values determined for the cube, are due to a  viable deformation
from the perfectly cubic shape. Taking the best-fitted results of
$\theta_{12}$ and $\theta_{23}$ as inputs, we determine the central
value of $\sin^22\theta_{13}$ should be $0.0238$, with a relatively
large error tolerance; this value lies in the range of measurement
precision of the Daya Bay experiment and is consistent with recent
results from the T2K Collaboration. \\

PACS: 14.60.Pq Neutrino mass and mixing

\end{abstract}

\draft

\maketitle

\section{Introduction}

Neutrino oscillation observations have revealed evidence that
neutrinos are massive. Neutrinos are produced via weak interaction
as flavor eigenstates $\nu_f=(\nu_e, \nu_\mu, \nu_\tau)$ and can be
written in the mass basis $\nu_m=(\nu_1, \nu_2, \nu_3)$, which are
really the physical states. These two bases are related by a unitary
matrix $U_\nu$, i.e., $\nu_f=U_\nu\nu_m$. The mixing in the lepton
sector is named as the Pontecorvo \cite{PMNS1}-Maki-Nakawaga-Sakata
\cite{PMNS2} (PMNS) matrix, which can account for the currently
available data on the observation of solar, atmospheric neutrino
oscillations and the reactor and accelerator neutrino experiments
\cite{PDG}. In the standard model,  the weak charged currents are
\begin{eqnarray}
\mathcal{J}^\mu=\bar{l}_i\gamma^\mu(1-\gamma_5)(U_l^\dag U_\nu)_{ij}
\nu_j, \label{CC2}
\end{eqnarray}
where $i,j=1, 2, 3$ and correspond to physical particles. The mixing
matrix
\begin{eqnarray}
U_{PMNS}=U_l^\dag U_\nu, \label{PMNS1}
\end{eqnarray}
is a $3\times3$ unitary matrix and can be parameterized by three
mixing angels $\theta_{12}$, $\theta_{23}$, and $\theta_{13}$, and
one $CP$ phase $\delta$ \cite{PDG},
\begin{eqnarray}
U_{PMNS}=\left(
\begin{array}{ccc}
c_{12}c_{13}&s_{12}c_{13}&s_{13}e^{-i\delta}\\
-s_{12}c_{23}-c_{12}s_{23}s_{13}e^{i\delta}&
c_{12}c_{23}-s_{12}s_{23}s_{13}e^{i\delta}&s_{23}c_{13}\\
s_{12}s_{23}-c_{12}c_{23}s_{13}e^{i\delta}&
-c_{12}s_{23}-s_{12}c_{23}s_{13}e^{i\delta}&c_{23}c_{13}
\end{array}
\right), \label{PMNS2}
\end{eqnarray}
where $c_{ij}\equiv\cos\theta_{ij}$, $s_{ij}\equiv\sin\theta_{ij}$.
If neutrinos are Majorana particles, there would be an additional
diagonal matrix diag$(e^{i\alpha_1/2}, e^{i\alpha_2/2}, 1)$
multiplied to the above $U_{PMNS}$ matrix, which is not relevant for
neutrino oscillations. The parametrization Eq. (\ref{PMNS2}) can be
rewritten as a product of three rotations $R_{ij}$ in the $ij$ plane
through angles $\theta_{ij}$ and a diagonal $CP$ phase matrix
$U_\delta$=diag$(e^{i\delta/2}, 1, e^{-i\delta/2})$,
\begin{eqnarray}
U_{PMNS}=R_{23}(\theta_{23})U_\delta^\dag
R_{13}(\theta_{13})U_\delta R_{12}(\theta_{12}),\label{PMNS3}
\end{eqnarray}
with
\begin{eqnarray}
R_{23}=\left(\begin{array}{ccc}1&0&0\\
0&c_{23}&s_{23}\\0&-s_{23}&c_{23}
\end{array}\right),
R_{13}=\left(\begin{array}{ccc}c_{13}&0&s_{13}\\
0&1&0\\-s_{13}&0&c_{13}\end{array}\right),
R_{12}=\left(\begin{array}{ccc}c_{12}&s_{12}&0\\
-s_{12}&c_{12}&0\\0&0&1\end{array}\right). \label{PMNS4}
\end{eqnarray}

There have been numerous phenomenological Ans\"{a}tze for the
entries of $U_{PMNS}$, for example, the democratic
\cite{democratic}, the bimaximal \cite{bimaximal}, and the
tribimaximal Ans\"{a}tze \cite{tribimaximal}. Among them, the
tribimaximal mixing is closer to the experimentally observed mixing
patterns, and the matrix is given by
\begin{eqnarray}
U_{tribi}=\left(\begin{array}{ccc}2/\sqrt{6}&1/\sqrt{3}&0\\
-1/\sqrt{6}&1/\sqrt{3}&1/\sqrt{2}\\
1/\sqrt{6}&-1/\sqrt{3}&1/\sqrt{2}\end{array}\right),\label{tribi}
\end{eqnarray}
which suggests $\theta_{12}$ = $\sin^{-1}(1/\sqrt{3})$,
$\theta_{23}$ = $\pi/4$, $\theta_{13}$ = 0. As noted, the $CP$ phase
$e^{i\delta}$ is always associated with $s_{13}$ [Eq.
(\ref{PMNS2})]; thus, null $\theta_{13}$ would imply that one cannot
observe $CP$ violation at lepton sector in the framework of the
standard model even though $\delta$ is not zero. Obviously, there is
no priori that the $CP$ violation should appear at the lepton
sector, but only nonzero $\theta_{13}$ can intrigue an enthusiasm to
explore $CP$ violation at the lepton sector. Once the $\theta_{13}$
is determined to be nonzero as the T2K experiment and our
theoretical prediction made in this work suggest, the next step
would be searching for $CP$ violation at the lepton sector.

Indeed, the tribimaximal symmetry  is well manifested by the data. A
rigorous symmetry would demand $\theta_{13}$ to be zero; however, it
is not the whole story because this elegant symmetry is to be
broken, and a nonzero $\theta_{13}$ is expected. The question is if
it is not zero, what is its size, which is the main concern of the
recent studies.

\begin{figure}[t]
\centering
\includegraphics[height=8.5cm, width=10.5cm, angle=0]{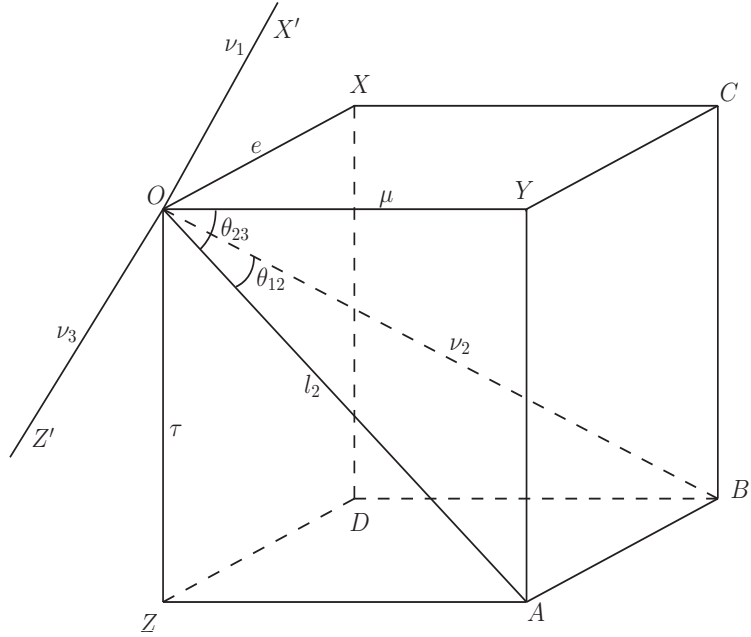}
\caption{Geometric representation of the tribimaximal mixing in Eq.
(\ref{tribi}) and (\ref{Lee}).} \label{FigCube}
\end{figure}

The unbroken tribimaximal matrix Eq. (\ref{tribi}) can  be further
written as a sequential product of two independent rotations on 12
and 23 planes:
\begin{eqnarray}
R_{23}(\pi/4)=\left(\begin{array}{ccc}1&0&0\\
0&1/\sqrt{2}&1/\sqrt{2}\\
0&-1/\sqrt{2}&1/\sqrt{2}\end{array}\right),
R_{12}(\sin^{-1}(1/\sqrt{3}))=\left(\begin{array}{ccc}2/\sqrt{6}&1/\sqrt{3}&0\\
-1/\sqrt{3}&2/\sqrt{6}&0\\
0&0&1\end{array}\right), \label{R23R12}
\end{eqnarray}
and $R_{13}$ becomes a $3\times 3$ unit matrix, i.e.,
$U_{tribi}=R_{23}(\pi/4)R_{12}(\sin^{-1}(1/\sqrt{3}))$. Friedberg
and Lee \cite{Lee1} propose a geometrical interpretation for the
tribimaximal symmetry as shown in Fig. \ref{FigCube}. For readers'
convenience, let us briefly introduce Friedberg and Lee's
geometrical model and their conventions \cite{Lee1}. The charged
leptons in the basis $L_c=(S_e,\; S_\mu,\; S_\tau)^T$ correspond to
the three mutually perpendicular sides of a cube, and the neutrino
basis $L_n=(S_{\nu_1},\;S_{\nu_2},\;S_{\nu_3})^T$ corresponds to
another coordinate system (see Fig. \ref{FigCube}). These two
coordinate systems are related to each other by rotations. One can
perform two independent rotations to associate them.
$R_{23}^\dag(\pi/4)$ and $R_{12}(\sin^{-1}(1/\sqrt{3}))$ transform
the two independent bases into a common one. These practical
operations are described below. $R_{23}^\dag(\pi/4)$ mixes the
second and third components of the basis $L_c$ and keeps the first
one invariant to get a new basis $(S_1,\; S_2,\; S_3)^T$, while
$R_{12}(\sin^{-1}(1/\sqrt{3}))$ mixes the first and second
components of $L_n$, retaining the third one invariant to reach the
same basis $(S_1,\; S_2,\; S_3)^T$. The mathematical expressions for
relating $(S_1, \;S_2, \;S_3)^T$ with the charged lepton basis
$(S_e, \;S_\nu, \;S_\tau)^T$ and neutrino basis $(S_{\nu_1}, \;
S_{\nu_2}, \;S_{\nu_3})^T$ are shown as follows:
\begin{eqnarray}
\left(\begin{array}{c}S_1\\ S_2\\
S_3\end{array}\right)=R_{23}^\dag(\pi/4) \left(\begin{array}{c}S_e\\
S_\mu\\ S_\tau\end{array}\right), \left(\begin{array}{c}S_1\\ S_2\\
S_3\end{array}\right)=R_{12}(\sin^{-1}(1/\sqrt{3}))\left(\begin{array}{c}S_{\nu_1}\\
S_{\nu_2}\\ S_{\nu_3}\end{array}\right). \label{Lee}
\end{eqnarray}
Following the convention given in Ref. \cite{Lee1}, when we discuss
the geometry structure, we abbreviate the sides $S_{l}$
($S_{\nu_i}$) as $l$ $(\nu_i)$ without causing any confusion. Here
$S_{e,\mu,\tau}$ and $S_{\nu_1,\nu_2,\nu_3}$ just refer to the
corresponding geometrical quantities marked in Fig. \ref{FigCube}
and are by no means the physical states.

Comparing $U_{tribi}=R_{23}(\pi/4)R_{12}(\sin^{-1}(1/\sqrt{3}))$
with $U_{PMNS}=U_l^\dag U_\nu$, it appears that the two rotations
$R_{23}^\dag(\pi/4)$ and $R_{12}(\sin^{-1}(1/\sqrt{3}))$ correspond
to the mixing matrices for the charged leptons and  neutrinos,
respectively. It is noted that in Eq. (\ref{Lee}), we only concern
the mixing parts; thus, inserting $\gamma^0\gamma^{\mu}(1-\gamma_5)$
between $((S_1,\;S_2,\;S_3)^T)^{\dagger}$ and $(S_1,\;S_2,\;S_3)^T$
which is irrelevant to our geometrical settings, we just derive the
Lagrangian of weak interaction. We also would like to point out that
in Eq. (\ref{Lee}), the high symmetry is assumed, and all quantities
are indeed corresponding to the zeroth order ones \cite{Lee1}, and
then later when we introduce a deformation of the cube to break the
tribimaximal symmetry, the concerned quantities would turn into the
physical ones.

Concretely, in Fig. \ref{FigCube}, sides $OX$, $OY$, and $OZ$
represent $e$, $\mu$, and $\tau$; and $\nu_1$, $\nu_2$, and $\nu_3$
correspond to $OX^\prime$, $OB$, and $OZ^\prime$, respectively. The
line-$OZ'$ resides on the plane $OZAY$ and spans an angle of $\pi/4$
with respect to the $OZ$ axis, whereas line $OX'$, line $OZ'$, and
line $OB$ compose three-dimensional mutually perpendicular
coordinate axes, and according the right-hand rule, we have an
$OB-OZ'-OX'$ system. The angle spanned between $OX'$ and $OX$ is
$\theta_{12}$. Therefore, the two rectangular coordinate systems
transform from each other by two rotations about the axes $OX$ and
$OZ'$, respectively. The right-handed rotation $R_{23}^\dag(\pi/4)$
about the $OX$ axis brings $\mu$ to $OA$ and $\tau$ to $\nu_3$, and
a second right-handed rotation $R_{12}(\sin^{-1}(1/\sqrt{3}))$, with
$\theta_{12}=\sin^{-1}(1/\sqrt{3})$, turns $\nu_1$ into $e$ and
$\nu_2$ into $OA$. Then, after performing the two successive
operations, the basis $(S_1, \;S_2, \;S_3)^T$ shown above can be
directly read out as $(e, \;OA, \;\nu_3)^T$.

Although the tribimaximal mixing Ansatz is close to the experimental
data and exhibits a striking symmetry, it is not the exact form of
the PMNS matrix. Moreover, this symmetry demands $\theta_{13}$ to be
zero. If the tribimaximal symmetry is not exact, with the angles
$\theta_{23}$ and $\theta_{12}$ obviously deviate from the values
determined by the symmetry, one has sufficient reason to believe
that $\theta_{13}$ should not be zero. In fact, the previous
measurements set a lower bound for $\theta_{13}$ as
$\sin^22\theta_{13}< 0.15$ \cite{PDG}, and will be more precisely
measured at the upcoming reactor experiments Daya Bay \cite{DayaBay}
and Double Chooz \cite{DoubleChooz}.

It would be interesting to investigate how to break the tribimaximal
symmetry from a theoretical aspect. Friedberg and Lee suggest to
break the symmetry from the charged lepton side \cite{Lee2}, whereas
He and his collaborators break the symmetry from the neutrino sector
\cite{HJHe}. Since the whole mixing matrix is a product of the two
unitary matrices that, respectively, diagonalize the charged lepton
and neutrino mass matrices as $U_{PMNS}=U_l^\dag U_\nu $, breaking
from either side is just like climbing up Mount Everest from the
south or north side as Lee comments\cite{workshop}. Their schemes to
break the symmetry are algebraic.

Instead, in this work, we propose to break the symmetry based on
Friedberg and Lee's geometrical picture. Namely, we let the cube  be
slightly deformed and the nonzero $\theta_{13}$ value would emerge.
Concretely, by deforming the geometric representation of the
tribimaximal mixing, the angles would deviate from their ideal
values; by fitting them to the data, we determine the deformation
scale of the cube, and then by the new geometric shape $\theta_{13}$
is no longer zero.

The work is organized as follows. In Sec. II, we present our
geometrical model of deforming the cube to get the $\theta_{13}$ as
a function of the other two mixing angles. Then, in Sec. III, we
present our numerical results.  The last section is devoted to our
conclusion and some discussions.


\begin{figure}[t]
\centering
\includegraphics[height=8.5cm, width=10.5cm, angle=0]{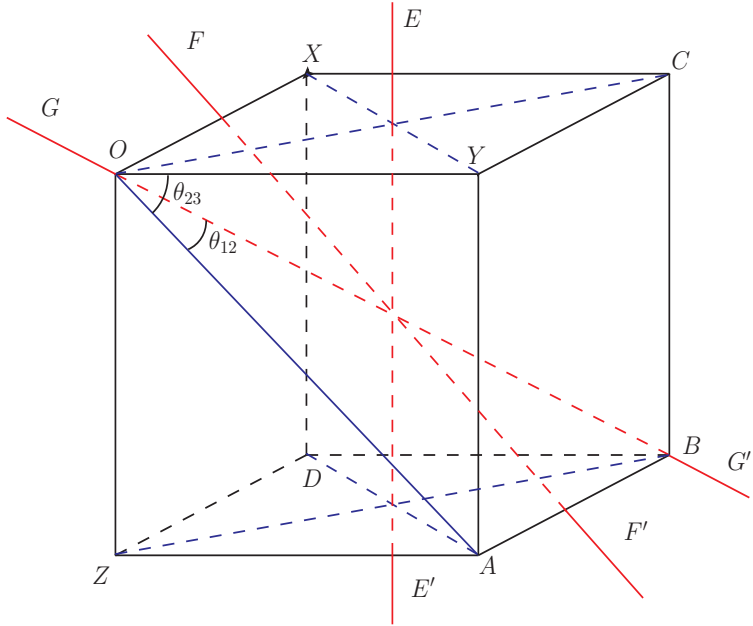}
\caption{(color online) The three classes of symmetry axes for a
cube, i.e., $EE^{\prime}$, $FF^{\prime}$, and $GG^{\prime}$.}
\label{CubeAxis}
\end{figure}

\section{The Deformed Cube Model}

It is noticeable that the angle between lines $OA$ and $OB$  and
that between $OY$ and $OA$ in the cube are precisely the two mixing
angles of the tribimaximal matrix $\theta_{12}$ and $\theta_{23}$,
respectively. For a perfect symmetry, which corresponds to a
complete cube, we have $\theta_{12}=\sin^{-1}(1/\sqrt{3})$ and
$\theta_{23}={\pi/4}$, which are determined by the geometry. It is
then viable that a deformation would lead to the more realistic form
of the PMNS matrix, and, thus, the nonzero $\theta_{13}$ would
emerge. After this deformation, $\theta_{12}$ and $\theta_{23}$ are
not the values given above anymore, but dependent on the form of the
deformation. A cube is a kind of polyhedron with high symmetry
described by a certain group, so a deformation of a cube should be
regarded as a symmetry breaking.

Now, let us demonstrate how to deform the above cube. For choosing
the deformation scheme, we set three principles:
\begin{itemize}
\item There are three rotation axes for a cube as presented
in Fig. \ref{CubeAxis}, i.e., $EE^{\prime}$, $FF^{\prime}$, and
$GG^{\prime}$. Apparently, the axis $GG^{\prime}$ is related with
the mixing angle $\theta_{12}$ ($\angle AOB$), and the axis
$FF^{\prime}$, which is parallel to the side OA, is related to
$\theta_{23}$ ($\angle AOY$). Then, the rest symmetry axis
$EE^{\prime}$ may be related to the zero $\theta_{13}$ in the
tribimaximal mixing. Thus, after the supposed deformation, the three
symmetries would be broken, and the value of the deformation angle
is related to $\theta_{13}$. For simplicity, we just choose the
deformation angle to be $\theta_{13}$.
\item For the tribimaximal mixing, $\theta_{23}$ = $\pi/4$ and
$\theta_{13}$ = 0, there exists the $\mu - \tau$ symmetry
\cite{mutau}. The global fit \cite{fit} gives $\theta_{23} =
42.8^\circ$, which apparently breaks the $\mu-\tau$ symmetry. Thus,
in the deformation, $\theta_{23}$ ($\angle AOY$) should be changed
from $\pi/4$ to some values in order to break the $\mu-\tau$
symmetry.
\item In Ref. \cite{fit}, the global fits of $\theta_{12}$ and
$\theta_{23}$ are $34.4^\circ$ and $42.8^\circ$, respectively. Thus,
for the deformation, $\theta_{12}$ ($\angle AOB$) and $\theta_{23}$
($\angle AOY$) should be changed toward smaller values than
$\sin^{-1}(1/\sqrt{3})$ and $\pi/4$, respectively.
\end{itemize}

\begin{figure}[t]
\centering
\includegraphics[height=8.5cm, width=10.5cm, angle=0]{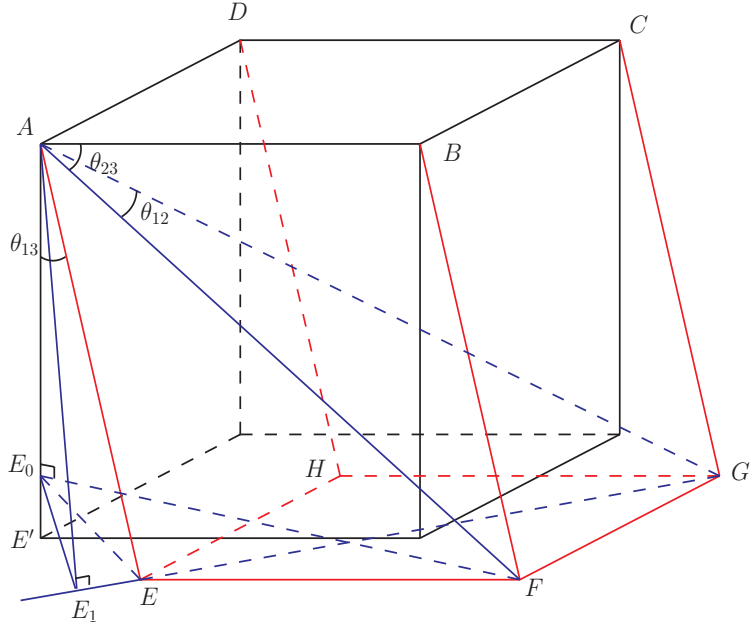}
\caption{(color online) The sketch of the shift of the bottom face
relative to the top face.} \label{twist}
\end{figure}

Considering above principles, the simplest and most direct scheme to
deform the cube is to  slide the bottom face parallel to the top
face, and a small angle would emerge, and this angle is identified
as $\theta_{13}$. The length of each side is unchanged during the
slide. This operation is explicitly illustrated in Fig. \ref{twist}.
With the parallel slide, the bottom face becomes $EFGH$. To be
consistent with Friedberg and Lee's picture and the principles we
proposed above, we identify $\theta_{12}$ = $\angle FAG$,
$\theta_{23}$ = $\angle BAF$, and $\theta_{13}$ = $\angle EAE_0$.
$E_0$ is the point of intersection between side $AE^{\prime}$ and
plane $EFGH$. $AE_1$ and $E_0E_1$ are perpendicular to the diagonal
line $EG$.

Setting $\angle E_0EG \equiv \alpha$ and in  the rectangular
triangle Rt$\bigtriangleup AE_1G$, one has
\begin{eqnarray}
AG^2=AE_1^2+E_1G^2=\cos^2\theta_{13}+\sin^2\theta_{13}\sin^2\alpha
+(\sqrt{2}+\sin\theta_{13}\cos\alpha)^2. \label{AG1}
\end{eqnarray}
In $\bigtriangleup AEF$,
\begin{eqnarray}
AG^2=1+4\cos^2\theta_{23}-4\cos\theta_{23}\cos\left(\pi-\theta_{12}
-\sin^{-1}(2\sin\theta_{12}\cos\theta_{23})\right), \label{AG2}
\end{eqnarray}
and in $\bigtriangleup EE_0F$,
\begin{eqnarray}
E_0F^2=E_0E^2+EF^2-2E_0E\cdot EF \cos\angle E_0EF
\end{eqnarray}
\begin{eqnarray}
4\cos^2\theta_{23}-\cos^2\theta_{13}=\sin^2\theta_{13}+1
-2\sin\theta_{13}\cos(\alpha+{\pi\over 4}). \label{EE0F}
\end{eqnarray}

The geometrical relationship of the sides and angles in the deformed
cube would determine Eq. (\ref{AG1}), Eq. (\ref{AG2}), and Eq.
(\ref{EE0F}). From these equations, we can get an analytical
expression of $\theta_{13}$, with respect to the other two mixing
angles $\theta_{12}$ and $\theta_{23}$ as
\begin{eqnarray}
\sin^2\theta_{13}=4\cos^4\theta_{23}-4\cos^2\theta_{23}
+4\cos^2\theta_{23}\cos^2\left(\theta_{12}
+\sin^{-1}(2\sin\theta_{12}\cos\theta_{23})\right)+1. \label{sin13}
\end{eqnarray}

As we have mentioned above, a cube is a highly symmetric polyhedron
that could be represented by a global $S4$ group \cite{group}. This
group has 24 elements classified in five conjugate classes. As shown
in Fig. \ref{CubeAxis}, a cube has three kinds of rotation axes, $h
= 2$, $h=3$, and $h=4$, corresponding to $FF^\prime$, $GG^\prime$,
and $EE^\prime$, respectively. All the rotation axes in the same $h$
are equivalent.

It is notable that the angle between the axes of $h=2$ and $h=4$ is
$\pi/4$, and the angle between the axes of $h=2$ and $h=3$ is
$\sin^{-1}(\sqrt{1/3})$. In other words, the two angles are exactly
that in the tribimaximal form of the PMNS matrix. Another angle
corresponding to $\theta_{13}$ must be an angle between $h=4$ and
$h=4$ itself, so $\theta_{13}=0$.

With the deformation, the symmetry of the cube is  broken,
and $\theta_{13}$ acquires a nonzero value. We can then view
$\theta_{13}$ as the parameter representing the deviation from the
cubic symmetry. It is then viable to define $\theta_{13}$ as the
angle between the "new" $h=4$ axis and the "old" one.







\begin{figure}[t]
\centering
\includegraphics[height=8.5cm, width=10.5cm, angle=0]{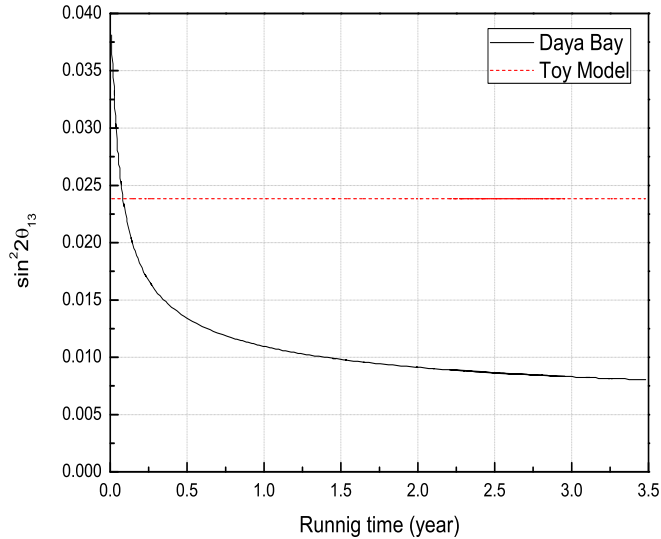}
\caption{(color online) Comparison of our result (dashed line) with
the Daya Bay expected sensitivity limit to $\sin^22\theta_{13}$ as a
function of running time.} \label{DayaBay}
\end{figure}
\begin{figure}[t]
\centering
\includegraphics[height=8.5cm, width=10.5cm, angle=0]{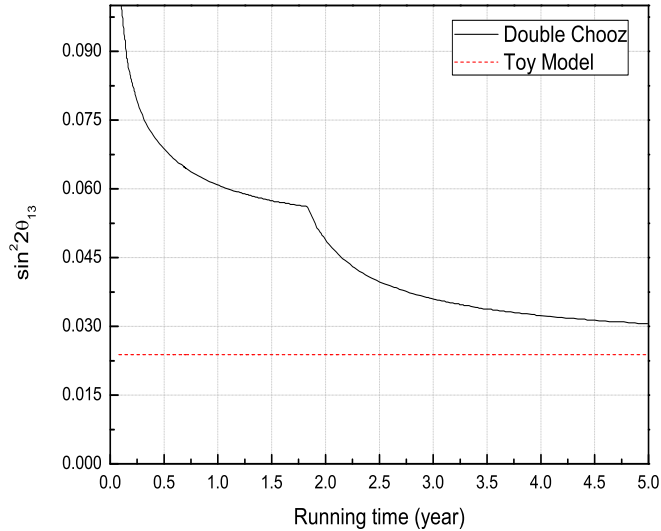}
\caption{(color online) Comparison of our result (dashed line) with
the Double Chooz expected sensitivity limit to $\sin^22\theta_{13}$
as a function of running time.} \label{DoubleChooz}
\end{figure}

\section{Numerical Results}

Ref. \cite{fit} presents the updated global fit to the
three-generation neutrino mixing:
\begin{eqnarray}
\theta_{12}=34.4\pm1.0^\circ,
\theta_{23}={42.8^{+4.7}_{-2.9}}^\circ. \label{fit}
\end{eqnarray}
Using the data as inputs, we obtain the numerical value of
$\theta_{13}$:
\begin{eqnarray}
\sin^22\theta_{13}=0.0238, \theta_{13}=4.44^\circ. \label{NuResult}
\end{eqnarray}
The errors of the fit would cause a theoretical uncertainty to
$\theta_{13}$:
\begin{eqnarray}
\sin^22\theta_{13}=0.0238^{+0.0762}_{-0.0238}.
\end{eqnarray}
The errors are rather large, and, in fact, to make
$\sin^22\theta_{13}\geq 0$, the lower bound shown in the above
expression is set. This expression indicates that our prediction on
$\theta_{13}$ is somehow sensitive to the input data and that in
order to get more precise values of $\theta_{13}$, more precise
values of the input are needed.

Two reactor neutrino experiments, Daya Bay \cite{DayaBay} and Double
Chooz \cite{DoubleChooz}, aiming to directly measure $\theta_{13}$
are expected to reach a very high precision. We illustrate a
relation  of the expected sensitivity of the Daya Bay and Double
Chooz  as a function of the running time in Figs. \ref{DayaBay} and
\ref{DoubleChooz}, respectively, where we mark the central value of
$\sin^22\theta_{13}$ calculated in this work.

Apparently, because of the high precision, the $\sin^22\theta_{13}$
value from our model would be probed at the first run of
the Daya Bay experiment.

\begin{figure}[t]
\centering
\includegraphics[height=8.5cm, width=10.5cm, angle=0]{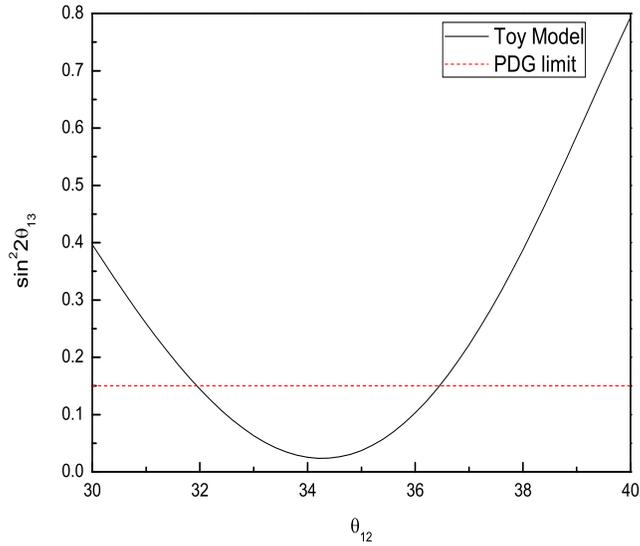}
\caption{(color online) $\sin^22\theta_{13}$ as a function of
$\theta_{12}$ for $\theta_{23}=42.8^\circ$ from the toy model (solid
line). The limit $\sin^22\theta_{13}<0.15$, CL=90$\%$ from PDG is
plotted as the dashed line.} \label{sin1223a}
\end{figure}
\begin{figure}[t]
\centering
\includegraphics[height=8.5cm, width=10.5cm, angle=0]{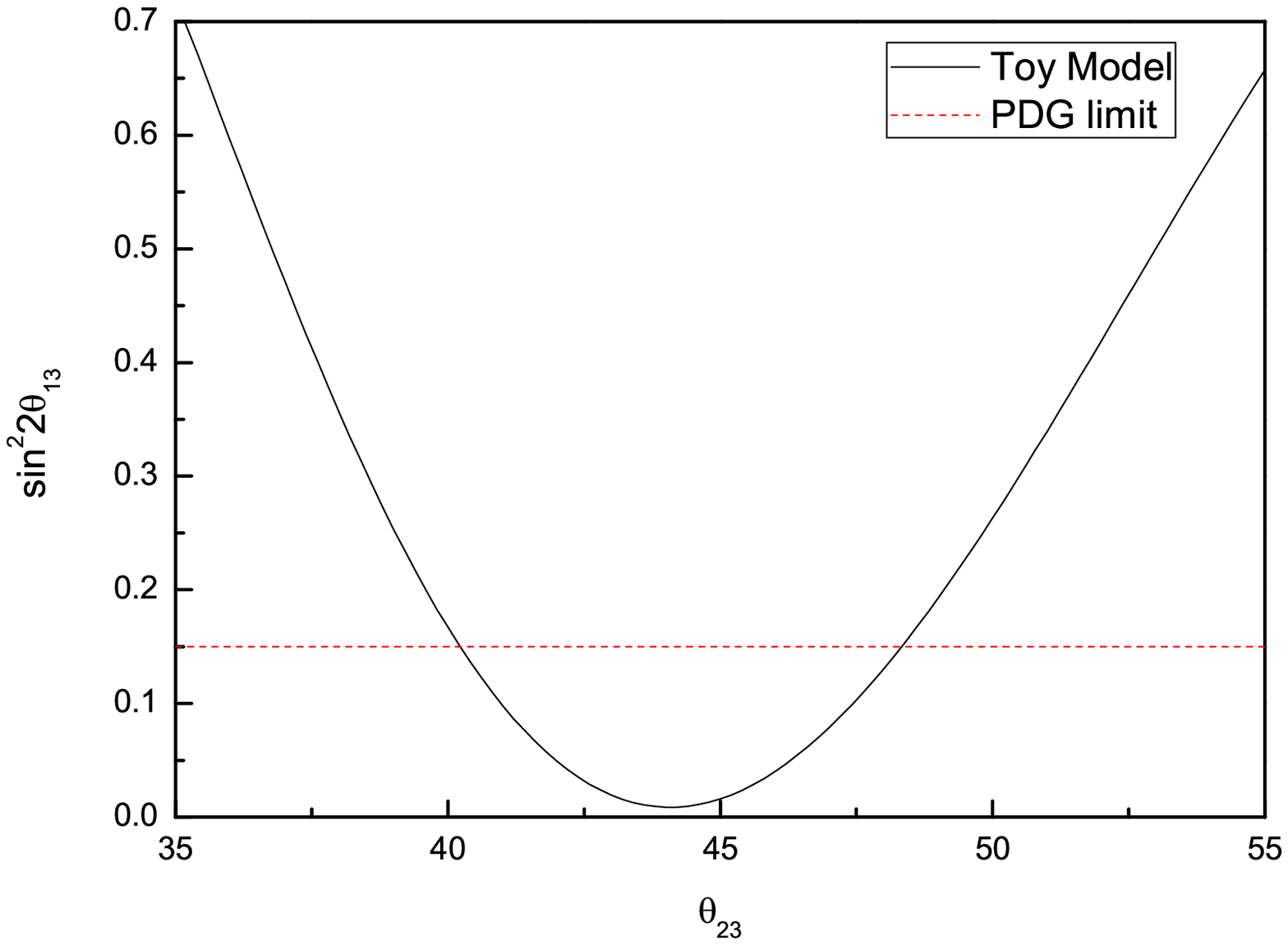}
\caption{(color online) $\sin^22\theta_{13}$ as a function of
$\theta_{23}$ for $\theta_{12}=34.4^\circ$ from the toy model (solid
line). The limit $\sin^22\theta_{13}<0.15$, CL=90$\%$ from PDG is
plotted as the dashed line.} \label{sin1223b}
\end{figure}

We show $\sin^22\theta_{13}$  as a function of $\theta_{12}$ for
$\theta_{23}=42.8^\circ$ in Fig. \ref{sin1223a}, and the dependence
on $\theta_{23}$ for $\theta_{12}=34.4^\circ$ in Fig.
\ref{sin1223b}. Particle Data Group (PDG) presents an upper bound of
$\theta_{13}$ as $\sin^22\theta_{13}<0.15$ at CL=$90\%$ \cite{PDG},
which is also marked in Figs. \ref{sin1223a} and \ref{sin1223b}.
From Figs. \ref{sin1223a} and \ref{sin1223b}, we can see that the
theoretically predicted value of $\sin^22\theta_{13}$ is sensitive
to both $\theta_{12}$ and $\theta_{23}$. By the updated data,
$\theta_{12}$ and $\theta_{23}$ are constrained within the range
($31.9^\circ$ $-$ $36.5^\circ$) and ($40.2^\circ$ $-$ $48.3^\circ$),
respectively.




\section{Discussions and conclusions}
$\theta_{23}={\pi / 4}$ and $\theta_{13}=0$ imply the so-called $\mu
- \tau$ symmetry \cite{mutau} embedding in the neutrino mass matrix,
i.e., the mass matrix in the flavor basis has an obvious $\nu_\mu -
\nu_\tau$ permutation symmetry. This leads to the mass matrix with
the form
\begin{eqnarray}
M=\left(\begin{array}{ccc}A&B&B\\
B&C&D\\ B&D&C\end{array}\right).
\end{eqnarray}
In Ref. \cite{HJHe}, the authors discussed the soft breaking of the
$\mu - \tau$ symmetry that arises from the Majorana mass term of the
heavy right-handed neutrinos in the minimal seesaw model. From their
$\mu - \tau$ symmetry breaking model, they derived a relation among
the mixing angles and Dirac $CP$ phase
\begin{eqnarray}
\theta_{23}-{\pi\over 4}=-\theta_{13} \cot\theta_{12} \cos\delta.
\label{HJHe}
\end{eqnarray}
For the case that the Dirac $CP$ phase $\delta = 0$ and substituting
the experimental fits $\theta_{12} = 34.4^\circ$ and $\theta_{23} =
42.8^\circ$ into Eq. (\ref{HJHe}), we obtain the  value of
$\theta_{13}$ as
\begin{eqnarray}
\sin^22\theta_{13}=0.00276, \theta_{13}=1.51^\circ. \label{He13}
\end{eqnarray}

Instead, in a parallel work, Friedberg and Lee\cite{Lee2} suggested
that one can break the $\mu-\tau$ symmetry at the charged lepton
side in terms of a perturbation method, and they also showed that
the breaking may lead to a nonzero $\theta_{13}$.

In this work, by deforming the cube that corresponds to a full
tribimaximal form of the mixing matrix according the proposed
principles, we derive the analytic relation among the three lepton
mixing angles, and, taking the experimental data as inputs, we
deduce the value of unknown mixing angle $\theta_{13}$. The result
gives $\sin^22\theta_{13}=0.0238$, i.e., $\theta_{13}=4.44^\circ$.
As noticed, our theoretical prediction favors smaller $\theta_{13}$.
As $\theta_{13}$ is to be measured at the Double Chooz and Daya Bay
experiments, our result indicates that in the future, there would be
a great opportunity to fix the mysterious $\theta_{13}$. The recent
measurement of the T2K collaboration \cite{T2K} indicates that
$\sin^22\theta_{13}$ falls in a rather wide range of
$0.03(0.04)<\sin^22\theta_{13}<0.28(0.34)$, and the central value of
our theoretical prediction is consistent with the lower bound set by
the collaboration, and the error range is comparable. This value
also does not contradict the new measurement by Main Injector
Neutrino Oscillation Search \cite{MINOS}.

\begin{acknowledgments}
We thank Dr. Ye Xu for helpful discussion on the experimental
issues. This work is supported by the National Natural Science
Foundation of China, under Contract No. 11075079.

\end{acknowledgments}


\begin{thebibliography}{99}

\bibitem{PMNS1}
B. Pontecorvo, Zh. Eksp. Teor. Fiz. {\bf 33}, 549 (1957); {\bf 34},
247 (1958) [Sov. Phys. JETP {\bf 6}, 429 (1957)].

\bibitem{PMNS2}
Z. Maki, M. Nakagawa and S. Sakata, Prog. Theor. Phys. {\bf 28}, 870
(1962).

\bibitem{PDG}
K. Nakamura \textit{et al.} (Particle Data Group), J. Phys. G {\bf
37}, 075021 (2010).

\bibitem{democratic}
H. Fritzsch and Z.Z. Xing, Phys. Lett. B {\bf 372}, 265 (1996);
Phys. Lett. B {\bf 440}, 313 (1998); Phys. Rev. D {\bf 61} 073016,
(2000).

\bibitem{bimaximal}
F. Vissani, arXiv:hep-ph/9708483; V. D. Barger, S. Pakvasa, T.J.
Weiler and K. Whisnant, Phys. Lett. B {\bf 437}, 107 (1998); A.J.
Baltz, A.S. Goldhaber and M. Goldhaber, Phys. Rev. Lett. {\bf 81},
5730 (1998); I. Stancu and D.V. Ahluwalia, Phys. Lett. B {\bf 460},
431 (1999); H. Georgi and S.L. Glashow, Phys. Rev. D {\bf 61},
097301 (2000).

\bibitem{tribimaximal}
P.F. Harrison, D.H. Perkins and W.G. Scott, Phys. Lett. B {\bf 530},
167 (2002); Z.Z. Xing, Phys. Lett. B {\bf 533}, 85 (2002); P.F.
Harrison and W.G. Scott, Phys. Lett. B {\bf 535}, 163 (2002); X.G.
He and A. Zee, Phys. Lett. B {\bf 560}, 87 (2003); I. Stancu and
D.V. Ahluwalia, Phys. Lett. B {\bf 460}, 431 (1999).

\bibitem{Lee1}
R. Friedberg and T.D. Lee, Annals Phys. (N.Y.) {\bf 324}, 2196
(2009).

\bibitem{DayaBay}
J. Cao, Nucl. Phys. B, Proc. Suppl. {\bf 155}, 229 (2006); S.M.
Chen, J. Phys. Conf. Ser. {\bf 120}, 052024 (2008); C. White, J.
Phys. Conf. Ser. {\bf 136}, 022012 (2008).

\bibitem{DoubleChooz}
C. Palomares (Double Chooz Collaboration), Proc. Sci., EPS-HEP2009
(2009) 275; F. Ardellier \textit{et al.}, arXiv:hep-ex/0606025;
arXiv:hep-ex/0405032.

\bibitem{Lee2}
R. Friedberg and T.D. Lee, Chinese Phys. C {\bf 34}, 1547 (2010).

\bibitem{HJHe}
S. F. Ge, H. J. He, and F. R. Yin, J. Cosmol. Astropart. Phys. 05
(2010) 017; H. J. He and F.R. Yin, Phys. Rev. D {\bf 84}, 033009
(2011).

\bibitem{workshop}
T.D. Lee, Proceedings for the Conference on the Neutrino Physics in
the Daya Bay Era, Beijing, China, 2010 (unpublished).


\bibitem{mutau}
R.N. Mohapatra and A.Yu. Smirnov, Ann. Rev. Nucl. Part. Sci. {\bf
56}, 569 (2006).

\bibitem{fit}
M. C. Gonzalez-Garcia, M. Maltoni, and J. Salvado, J. High Energy
Phys., 04 (2010) 056.

\bibitem{group}
H. Ishimori, T. Kobayashi, H. Ohki, H. Okada, Y. Shimizu, and M.
Tanimoto, Prog. Theor. Phys. Suppl. {\bf 183}, 1 (2010).




\bibitem{T2K}
K. Abe \textit{et al.} (T2K Collaboration), Phys. Rev. Lett. {\bf
107}, 041801 (2011).

\bibitem{MINOS}
P. Adamson \textit{et al.}, (MINOS Collaboration),
arXiv:hep-ex/1108.0015.

\end{thebibliography}
\end{document}